\documentclass[11pt]{article}
\usepackage{amssymb}
\usepackage{citesort}
\begin{document}
\begin{titlepage}

\begin{flushright}
hep-th/0604187\\
BONN--TH--2006--004 \\
 IFT UWr 0104--006
\end{flushright}
\bigskip

\begin{center}
{\LARGE\bf\sf Liouville theory
and
uniformization \\
\vskip 2mm
 of four-punctured sphere}
\end{center}

\bigskip

\begin{center}
    {\large\bf\sf
    Leszek Hadasz}\footnote{Alexander von Humboldt Fellow;
    e-mail: hadasz@physik.uni-bonn.de} \\
\vskip 1mm
    Physikalisches Institut\\
    Rheinische Friedrich-Wilhelms-Universit\"at \\
    Nu\ss{}allee 12, 53115 Bonn, Germany
    \\
    and \\
    M. Smoluchowski Institute of Physics, \\
    Jagiellonian University, \\
    Reymonta 4, 30-059 Krak\'ow, Poland \\
\vskip 5mm
    {\large\bf\sf
    Zbigniew Jask\'{o}lski}\footnote{e-mail: jask@ift.uni.wroc.pl
}\\

\vskip 1mm
   Institute of Theoretical Physics\\
 University of Wroc{\l}aw\\
 pl. M. Borna, 50-204 Wroc{\l}aw,
 Poland
\end{center}
\vskip .5cm
\begin{abstract}
Few years ago Zamolodchikov and Zamolodchikov proposed an expression
for the 4-point classical Liouville action in terms of the 3-point
actions and the classical conformal block
\cite{Zamolodchikov:1995aa}. In this paper we develop a method of
calculating the uniformizing map and the uniformizing group from the
classical Liouville action on $n$-punctured sphere and discuss the
consequences of Zamolodchikovs conjecture for an explicit
construction of the uniformizing map and the uniformizing group for
the sphere with four punctures.
\end{abstract}
\end{titlepage}
\section{Introduction}
\setcounter{equation}{0}

The uniformization problem for $n$-punctured $ ( n\geqslant 3 )$
spheres can be formulated as follows \cite{Hempel}:

\vskip 3mm

{\em
Given $ X = {\mathbb C}\setminus\{z_1,\ldots,z_{n-1}\}$,
find the} ({\em unique to within conjugacy}) {\em Fuchsian group
$G\subset {\rm PSU}(1,1)$ which makes $\Omega$ conformally equivalent to the quotient
$\Delta/G$ of the unit disc $\Delta=\{z\in \mathbb{C}:|z|<1\}$
by $G$.}

\vskip 3mm

The existence of solution to this problem, called the {\it uniformization theorem}, was first proved by Poincar\'e
and Koebe in 1907.
The universal covering map $\lambda : \Delta \to \Delta/G$
is however explicitly known only for the thrice-punctured sphere \cite{Ahlfors}
and in few very special, symmetric cases with higher number of punctures \cite{Hempel}.
In particular an explicit construction of this map for the four-punctured sphere
is a long standing and still open problem.

One possible approach, going back to Poincar\'e, is based on the relation
of the uniformization problem to
a certain Fuchs  equation on $X$.
If  $\lambda : \Delta \to \Delta/G\cong X$ is the universal covering map,
the  inverse   $\rho=\lambda^{-1} :X \to \Delta  $
is a multi-valued function with branching points $z_j$ and with branches related
by elements of the covering group $G\subset$ PSU(1,1).
One can show that the  Schwarzian derivative of $\rho$
is a holomorphic function on $X$ of the form \cite{Hempel}
\begin{eqnarray}
\label{schwarz}
\{\rho,z\}
& = &
\frac12\sum_{k=1}^{n-1}\ \frac{1}{(z-z_k)^2}
+
\sum_{k=1}^{n-1}\ \frac{2 c_k}{z-z_k}\,,
\\
\label{schwarz:infty}
\{\rho,z\}
& \stackrel{z \to \infty}{=} &
\frac{1}{2z^2} + {\cal O}\left(z^{-3}\right),
\end{eqnarray}
where the {\it accessory parameters} $c_j$ satisfy the relations
\begin{equation}
\label{c_rel}
\sum_{k=1}^{n-1}\ c_k \; = \;  0,
\hskip 1cm
\sum_{k=1}^{n-1}\ (4c_k z_k + 1) \; = \; 1.
\end{equation}

It is a well
 known property of the Schwarzian derivative \cite{Hempel,Takhtajan:2001uj}
 that the map $\rho$
is, to within a M\"{o}bius transformation, a
quotient  of two linearly independent solutions
of the Fuchs equation
\begin{equation}
\label{fuchs}
\partial_z^2 \Psi +{1\over 2}\{\rho,z\}\Psi = 0.
\end{equation}
This in particular means that
there exists a unique to within SU(1,1) transformation
fundamental system $\{\Psi_1, \Psi_2\}$
of normalized (i.e.\ with the Wronskian equal to 1) solutions for which
\begin{equation}
\label{system}
\rho = {\Psi_1\over \Psi_2}.
\end{equation}
Note that any such system has to have an SU(1,1) monodromy with respect to all
punctures.

Whether the Fuchs equation can be used to
calculate the map $\rho$ depends on our ability
to calculate the accessory parameters and
to choose an appropriate fundamental system of normalized
solutions.
The first problem can be easily solved for three punctures where the accessory
parameters are completely determined by relations (\ref{c_rel}). It
is however  difficult  and  still unsolved for $n>3$.

One can get some more insight by relating this problems to the Liouville equation
on $X$.
Since PSU(1,1) is the isometry group of the Poincar\'e hyperbolic metric
$
g_\Delta={4d\rho d\bar \rho\over ( 1-|\rho|^2)^2}
$ on $\Delta$
the pull back
\begin{equation}
\label{sol}
\rho^* g_\Delta = {4\over ( 1-|\rho|^2)^2}
\left|{\partial \rho\over \partial z}\right|^2 dz d\bar z
= {\rm e}^{\varphi(z,\bar z)} dz d\bar z
\end{equation}
is a regular hyperbolic metric on $X$, conformal to the standard flat metric
$
dz d\bar z
$
on $X\subset \mathbb{C}$.
Its conformal factor $\varphi$
satisfies
the Liouville equation
\begin{equation}
\label{Liouville}
\partial_z\partial_{\bar z} \varphi(z,\bar z)
\; = \;
\frac{1}{2} {\rm e}^{\varphi(z,\bar z)}
\end{equation}
and has the following asymptotic behavior at punctures
\begin{equation}
\label{asymptot:parabolic}
\varphi(z,\bar z) = \left\{
\begin{array}{lll}
-2\log |z- z_j |  -2\log \left|\log |z- z_j |\right| + O(1) & {\rm as } & z\to z_j, \\
-2\log |z| - 2\log \left|\log |z|\right| + O(1) & {\rm as } & z\to \infty.
\end{array}
\right.
\end{equation}

It is known that there exists a unique solution
to
(\ref{Liouville}) and (\ref{asymptot:parabolic})
\cite{Heins}.  One can show that the
{\em energy--momentum tensor} $T(z)$ of this solution is equal to one half of the Schwarzian derivative
(\ref{schwarz}):
\begin{equation}
\label{tensor}
T(z)\equiv
-{1\over 4} (\partial \varphi)^2
+{1\over 2} \partial^2\varphi
 = \;
\frac12\{\rho,z\}.
\end{equation}
This allows to calculate all accessory parameters once
the classical solution $\varphi$ is known.

The problem of selecting an
appropriate fundamental system is slightly less demanding.
As we shall see it is sufficient to know the next to the first two leading terms
of the asymptotic of the classical solution $\varphi$ at one arbitrary puncture.

Unfortunately, the  problem to find solutions to the Liouville equation
seems to be at least as hard as the problem
of calculating the map $\rho$ itself, and the reformulation does not help much on this
stage. In this framework however one can consider
a  more general problem of spheres with $n$ elliptic singularities characterized
by real parameters $0 < \xi_j < 1$.
Instead of asymptotic conditions (\ref{asymptot:parabolic})
we impose
\begin{equation}
\label{asymptot:elliptic}
\varphi(z,\bar z) = \left\{
\begin{array}{lll}
-2\left(1-\xi_j \right)\log | z- z_j |  + O(1) & {\rm as } & z\to z_j, \\
-2(1+\xi_n)\log | z- z_j | + O(1) & {\rm as } & z\to \infty.
\end{array}
\right.
\end{equation}
The existence and uniqueness of $\varphi$
was in this case proved by Picard \cite{Picard1,Picard2} (see
also \cite{Troyanov}). The solution can be
interpreted as a conformal factor of the complete, hyperbolic
metric with the Gaussian curvature $R = -1$ on \( X = {\mathbb
C}\setminus\{z_1,\ldots,z_{n-1}\} \) and conical singularities of
the opening angles $0<2\pi\xi_j<2\pi$ at the points $z_j.$
The $n$-punctured sphere discussed so far
corresponds to the limiting case $\xi_j\to 0$ for all $j=1,\dots,n$.

The notion of accessory parameters can be introduced in terms of
 the energy momentum tensor of the solution $\varphi$. In the present case it takes the form
\cite{Takhtajan:2001uj}:
\begin{eqnarray}
\label{T:classical}
T(z)&=&
\sum\limits_{j=1}^{n-1}
\left[\frac{\delta_j}{(z-z_j)^2} + \frac{c_j}{z-z_j}\right],
\\
\nonumber
T(z) & \stackrel{z \to \infty}{=} &\; \frac{\delta_n}{z^2} + {\cal O}\left(z^{-3}\right),
\end{eqnarray}
where
$
\delta_j = \frac{1-\xi_j^2}{4},\ j=1,\dots,n
$
are {\it classical conformal weights}. The multi-valued function $\rho:X\to \Delta$
is still of interest although it is not longer an inverse to the universal covering
of $X$. It can be used as before in formula (\ref{sol})
to construct solutions to the Liouville equation with asymptotic behavior
(\ref{asymptot:elliptic}).

The ideas described above are classic and most of them were already pursued by Poincar\'e.
For almost a century the problem of accessory parameters had reminded unsolved.
An essentially new insight was brought in
by the so called
 Polyakov conjecture in 1982 \cite{Polyakov82}.
 It states that the (properly defined and normalized) Liouville
action functional evaluated on the classical solution $\varphi_{\rm cl}(z,\bar z)$
is a generating function for the accessory parameters:
\begin{equation}
\label{generating} c_j = - \frac{\partial {S}^{\rm
(cl)}(\delta_i\,;\,z_i)}{\partial z_j} \ .
\end{equation}
This formula was derived within path integral approach to the quantum Liouville theory as
the quasi-classical limit of the conformal Ward identity \cite{Takhtajan:1994vt}.
In the case of the parabolic singularities on $n$-punctured Riemann sphere a rigorous proof based
on the theory of quasiconformal mappings was given by Zograf and Takhtajan \cite{ZoTa1}.
Other  proofs, valid both in the case of parabolic and general elliptic
singularities, were proposed in
\cite{Cantini:2001wr} and  \cite{Takhtajan:2001uj}.

The next significant step was done by Zamolodchikov and
Zamolodchikov \cite{Zamolodchikov:1995aa}. Analyzing the classical
limit of the four-point function of the quantum Liouville theory
they argued that the classical Liouville action for four elliptic
(parabolic) singularities can be expressed in terms of the
classical Liouville action for three singularities and some
special function called the {\it classical conformal block}.
Recently this conjecture has been successfully  tested by symbolic
and numerical calculations in \cite{Hadasz:2005gk}. It should be
stressed however that it is still  far from being rigorously
proved. The basic problem is the classical block itself which is
so far   accessible only via term by term calculation of the
classical limit of the quantum conformal block.

The aim of the present paper is to analyze to what extend the Zamolodchikovs conjecture
can provide an explicit construction of the uniformization
of four-punctured sphere.
Our motivation is to get a better insight into a geometric content of
this conjecture and to develop a theoretical framework
for its new numerical tests.
The results indicate that the classical conformal block plays
a central role in the problem and certainly deserves further investigations.

The content of the paper is as follows. In Section 2  we analyze the problem of selecting an
appropriate pair of solutions to the Fuchs equation in the
case of elliptic weights.
It is shown that all the information required is encoded in the
derivatives of
the classical Liouville action with respect to the parameters $\xi_j$.
The case of parabolic
singularities is obtained by taking an appropriate limit.
The main result is that in order to calculate the map $\rho$
from the Fuchs equation it is sufficient to know the classical Liouville
action as a function of $z_j$'s and $\xi_j$'s.

In Section 3 we analyze the problem of calculating monodromies of the Fuchs equation
once the accessory parameters are known. Only the case of 4 punctures is considered.
For the standard locations $0,x,1,\infty$ we develop  systematic expansions
of monodromy matrices at $0,x,1$ in terms of power expansions in $x$ and $1-x$.
The results of Sections 2 and 3 are general and are independent of the form in which
the classical Liouville action is available.

In section 4 the Za molodchikovs conjecture is formulated
and some schemes of calculation of the classical Liouville
action and accessory parameters are developed.
They lead to very efficient methods of numerical calculations.
It should be stressed however that many steps
in their derivation still require sound mathematical proofs.
Concluding this section we discuss some open problems and possible extensions
of our work.

\section{Map $\rho: X \to \Delta$}
\setcounter{equation}{0}

In the case of the sphere with $n$ elliptic singularities with weights
$
\delta_j = \frac{1-\xi_j^2}{4}
$
we define the functional
\begin{eqnarray*}
\nonumber
S_{\rm L}[\delta_i;\phi]
& = &
\frac{1}{4\pi}
\lim_{\epsilon\to 0}
S_{\rm L}^\epsilon[\delta_i;\phi], \\
S_{\rm L}^\epsilon[\delta_i;\phi]
& = &
\int\limits_{X_\epsilon}\!d^2z
\left[\left|\partial\phi\right|^2 + {\rm e}^{\phi}\right]
+  \sum\limits_{j=1}^{n-1}\left(1-\xi_j\right)
\hspace{-4mm}\int\limits_{|z-z_j|=\epsilon}\hspace{-4mm}|dz|\ \kappa_z \phi
+\left(1+\xi_n\right)
\hspace{-2mm}\int\limits_{|z|=\frac{1}{\epsilon}}\hspace{-2mm}|dz|\ \kappa_z \phi
\\
\nonumber
&&
- 2\pi\sum\limits_{j=1}^{n-1}\left(1-\xi_j\right)^2\log\epsilon
- 2\pi\left(1+\xi_n\right)^2\log\epsilon,
\end{eqnarray*}
where \( \textstyle X_\epsilon = {\mathbb
C}\setminus\left(\bigcup_{j=1}^n \{|z-z_j|< \epsilon\}
\cup \{|z|>\frac{1}{\epsilon}\}\right) \).

\noindent
The classical Liouville action ${\cal S}^{\rm (cl)}(\delta_i\,;\,z_i)$ is then defined as
\cite{Cantini:2001wr,Takhtajan:2001uj}
\begin{eqnarray}
\label{action}
\nonumber
{S}^{\rm (cl)}(\delta_i,z_i)
&=&
S_{\rm\scriptscriptstyle L}[\delta_i;\varphi_{\rm cl}],
\end{eqnarray}
where $\varphi_{\rm cl}(z,\bar z)$ is the unique solution to (\ref{Liouville})
and (\ref{asymptot:elliptic}).

Once the classical action is known one can use the Polyakov conjecture (\ref{generating})
to calculate the accessory parameters and write down the corresponding Fuchs
equation
\begin{equation}
\label{ellipticFuchs}
\partial_z^2 \Psi +T \Psi = 0,
\end{equation}
with the energy momentum tensor $T$ given by (\ref{T:classical}).
Our aim in this section is to select a fundamental system of solutions to this
equation such that their quotient yields a multi-valued function $\rho:X\to \Delta$ with
SU(1,1) monodromy at each  $z_j$ and regular for all $z\in X$.

To this end let us first observe that (\ref{Liouville})
and (\ref{tensor}) imply that
${\rm e}^{-\frac{\varphi(z,\bar z)}{2}}$
is a real solution to the Fuchs equations (\ref{ellipticFuchs}) and its complex conjugate.
It can be therefore expressed as a bilinear combination of any fundamental
system and its complex conjugate. In order to fix the freedom related to the
SU(1,1) transformations let us choose a normalized system with
diagonal monodromy at an arbitrarily chosen singular point $z_j$:
\begin{equation}
\label{psis}
\Psi_{\xi,\pm}^{(j)}(z) = \frac{A_j^{\pm 1}}{\sqrt\xi_j}
 (z-z_j)^{\frac{1\pm\xi_j}{2}}\left(1 + {\cal O}(z-z_j)\right),
 \hskip 1cm
 A_j\in\mathbb{R}.
\end{equation}

It follows from reality, positivity and single valuedness of
${\rm e}^{-\frac{\varphi(z,\bar z)}{2}}$ on $X$
that the parameter $A_j$ can be adjusted
such that
$$
{\rm e}^{-\frac{\varphi(z,\bar z)}{2}} = \frac{D_j}{2}
\left[\left|\Psi_{\xi,-}^{(j)}(z)\right|^2 - \left|\Psi_{\xi,+}^{(j)}(z)\right|^2\right],
$$
where $D_j$ is a positive constant. Although the formula above is
derived by considering a small neighborhood of $z_j,$ it holds for
all $z\in X$. Using it  one easily derive all the required
properties of the map
$$
\rho_j(z) ={\Psi_{\xi,+}^{(j)}(z)\over \Psi_{\xi,-}^{(j)}(z)}.
$$
One can in particular apply the formula (\ref{sol}) to construct the
hyperbolic metric with Gaussian curvature $-1$ on $X$. As this
metric coincides with ${\rm e}^{\varphi(z,\bar z)}dzd\bar z,$ the
constant $D_j$ has to be equal to 1 and
\begin{equation}
\label{solution}
{\rm e}^{-\frac{\varphi(z,\bar z)}{2}} = \frac{1}{2}
\left[\left|\Psi_{\xi,-}^{(j)}(z)\right|^2 - \left|\Psi_{\xi,+}^{(j)}(z)\right|^2\right].
\end{equation}
Analyzing the limit $z\to z_j$ one finds that $A_j^2={1\over 2 \xi_j}{\rm e}^{{1\over 2}f_j},$
where $f_j$ is the next to the leading term in the $\varphi_{\rm cl}(z,\bar z)$
asymptotic at $z\to z_j$.
On the other hand
\begin{equation}
\label{derivative}
\frac{\partial S^{\rm (cl)}(\delta_i\,;\,z_i)}{\partial \xi_j} =
\lim_{\epsilon\to 0}\Big[(1-\xi_j)\log\epsilon-
\frac{1}{4\pi}\int\limits_{|z-z_j|=\epsilon}\hspace{-4mm}|dz|\ \kappa_z \phi\Big] = {1\over 2}f_j,
\end{equation}
and consequently
\begin{equation}
\label{A:j}
A_j^2 = \frac{1}{2\xi_j}\exp
\left\{\frac{\partial}{\partial \xi_j}S^{\rm (cl)}(\delta_i\,;\,z_i)\right\}.
\end{equation}
This equation, along with the Polyakov conjecture, shows that that the classical Liouville
action contains all the information needed to calculate the map $\rho:X\to \Delta$ from the Fuchs equation.

To study the parabolic limit $\xi_j \to 0$ we shall
first define the pair $\tilde\Psi^{(j)}_{\xi,\pm}(z),$
related to $\Psi^{(j)}_{\xi,\pm}(z)$ through the SU(1,1) transformation,
\[
\left(\begin{array}{c}
\tilde\Psi_{\xi,+}^{(j)}(z) \\ \tilde\Psi_{\xi,-}^{(j)}(z)
\end{array}\right)
=
\frac12
\left(\begin{array}{cc}
\xi_j^{1/2} + \xi_j^{-1/2} & \xi_j^{1/2} - \xi_j^{-1/2} \\
\xi_j^{1/2} - \xi_j^{-1/2} & \xi_j^{1/2} + \xi_j^{-1/2}
\end{array}\right)
\left(\begin{array}{c}
\Psi_{\xi,+}^{(j)}(z) \\ \Psi_{\xi,-}^{(j)}(z)
\end{array}\right).
\]
From (\ref{A:j}) it can be shown in the general case (and will be illustrated by the explicit calculation
in the next section) that
\begin{equation}
\label{A:general}
A_j
\; = \; 1+ \xi_j a_j + {\cal O}(\xi_j^2)
\; = \;
{\rm e}^{\xi_j a_j} + {\cal O}(\xi_j^2),
\end{equation}
with $a_j$ independent of $\xi_j.$ From this fact and (\ref{psis}) it follows that the limit
\begin{equation}
\label{xi:to:0}
\left(\!\begin{array}{c}
\Psi_{+}^{(j)}(z) \\ \Psi_{-}^{(j)}(z)
\end{array}\!\right)
 =
\lim_{\xi_j\to 0}
\left(\!\begin{array}{c}
\tilde\Psi_{\xi,+}^{(j)}(z) \\ \tilde\Psi_{\xi,-}^{(j)}(z)
\end{array}\!\right)
\end{equation}
exists.

 The inverse of the map
\begin{equation}
\label{invuniform}
\rho^{(j)}(z) = \frac{\Psi_{-}^{(j)}(z)}{\Psi_{+}^{(j)}(z)}
\end{equation}
is the universal covering map of the punctured sphere $X$ by the Poincar\'e disc $\Delta$.
In particular, as
\[
\left|\lim_{z\to z_j} {\Psi^{(j)}_{-}(z) \over \Psi^{(j)}_{+}(z)}
\right| = 1,
\]
the puncture is mapped onto the point on the boundary of the Poincar\'e disc.

\section{Monodromy matrices}
\setcounter{equation}{0}

In this section we shall consider the Riemann sphere with four punctures at the  standard
locations
$z = 0, x, 1 $ and $\infty$.
In this case
\begin{eqnarray*}
T(z)
&  =  &
\frac{1}{4z^2} + \frac{1}{4(z-x)^2} + \frac{1}{4(z-1)^2}
+ \frac{c_1(x)}{z}
+ \frac{c_2(x)}{z-x}
+ \frac{c_3(x)}{z-1},
\end{eqnarray*}
and the relations (\ref{c_rel}) can be written in the form
\begin{equation}
\label{accessory}
c_1(x) \; = \;  \frac12 -  (1-x)c_2(x),
\hskip 1cm
c_3(x)  \; = \; -\frac12 - xc_2(x).
\end{equation}
Our aim is to calculate in this case the monodromies of the fundamental systems
of solutions to the Fuchs equation
(\ref{fuchs}) constructed in the previous section.

To this end  we shall split
$T(z)$ onto the ``free'' term $T_{0}(z)$ and the ``interaction'' $-V(z),$
\[
T(z) \; = \; T_0(z) - V(z).
\]
$T_0(z)$ contains by construction all terms in $T(z)$ singular at $z = 0,$ term with a
second order pole at $z=x$ plus a ``correction''
enforcing the behavior $\sim z^{-2}$ at the infinity,
\begin{eqnarray*}
T_0(z)
& = &
\frac{1}{4z^2} + \frac{1}{4(z-x)^2} + \frac{c_1(x)}{z} - \frac{c_1(x)}{z-x}
\; = \;
\frac{1}{4z^2} + \frac{1}{4(z-x)^2} - \frac{xc_1(x)}{z(z-x)} \\
& \equiv &
\frac{1}{4z^2} + \frac{1}{4(z-x)^2} - \frac{1+\nu^2(x)}{4z(z-x)}
\end{eqnarray*}
with
\begin{equation}
\label{nu}
\nu^2(x) = 4xc_1(x) - 1 = -4x(1-x)c_2(x) +x -(1-x),
\end{equation}
and
\[
V(z) \; = \; - \frac{(1-x)c_3(x)}{(z-x)(z-1)} -\frac{1}{4(z-1)^2}.
\]
Solutions to the ``free'' equation
\[
\partial^2_zf (z) + T_0(z)f(z) \; = \; 0
\]
are then expressible through the hypergeometric functions with the well-known
monodromies around $z = 0, x,$ while the solutions (and monodromies) of the
``full'' equation
\begin{equation}
\label{perturbative}
\partial^2_z\Psi(z) + T_0(z)\Psi(z) \; = \; V(z)\Psi(z)
\end{equation}
can be obtained by perturbation theory with a $n-$th order correction
proportional to $x^n.$

One then repeats the calculation above with $x \to 1-x.$ Since
this problem is related through the global conformal transformation $z \to w = 1-z$
to our original problem with the points $0$ and 1 exchanged (and
opposite orientation of the $z$ and $w$ planes) it is clear that the monodromy
matrix around $w=0$ yields the monodromy matrix around $z = 1.$
Note however that this time the parameter in the perturbative
expansion is $1-x$. Our method thus yields all three monodromy matrices\footnote{
The monodromy matrix around the puncture $z=\infty$ is equal to the inverse of the
product of the remaining three matrices.}
only for those $x$ for which both $x$ and $1-x$ are small enough.

Let
\begin{equation}
\label{T:0}
T_0^{(\xi,\epsilon)}(z) = \frac{1-\epsilon^2}{4 z^2}+ \frac{1-\xi^2}{4(z-x)^2} +
\frac{\xi^2+\epsilon^2 - \nu^2-1}{4z(z-x)}.
\end{equation}
We can choose the normalized basis in the space of solutions to the equation
\[
\partial^2f(z) + T_0^{(\xi,\epsilon)}(z) f(z) \; = \; 0
\]
in the form of the pair of functions with the diagonal monodromy matrix around $z=0,$
\begin{eqnarray*}
\phi^{(\xi,\epsilon)}_{\pm}(z)
&  = &
 {\sqrt\frac{x}{\epsilon}}\left(\frac{z}{x}\right)^{\frac{1\pm\epsilon}{2}}\,
 \left(1-\frac{z}{x}\right)^{\frac{1+\xi}{2}}
 {}_2F_1\!\left(\frac{1+ \xi + \nu \pm\epsilon}{2},
\frac{1 + \xi - \nu \pm\epsilon}{2},1\pm\epsilon,\frac{z}{x}\right),
\end{eqnarray*}
or in the form of the pair of functions with the diagonal monodromy matrix around $z=x,$
\begin{eqnarray*}
\psi^{(\xi,\epsilon)}_{\pm}(z)
&  = &
 {\rm e}^{\pm\xi a}{\sqrt\frac{x}{\xi}}\left(\frac{z}{x}\right)^{\frac{1+\epsilon}{2}}\,
 \left(1-\frac{z}{x}\right)^{\frac{1\pm\xi}{2}}
 {}_2F_1\!\left(\frac{1\pm\xi + \nu +\epsilon}{2},
\frac{1\pm\xi - \nu +\epsilon}{2},1\pm\xi,1-\frac{z}{x}\right).
\end{eqnarray*}
These two pairs of functions are obviously expressible through each other,
\[
\left(
\begin{array}{c}
\psi^{(\xi,\epsilon)}_+(z) \\
\psi^{(\xi,\epsilon)}_-(z)
\end{array}
\right)
\;  = \; {\cal C}^{(\xi,\epsilon)}\cdot
\left(
\begin{array}{c}
\phi^{(\xi,\epsilon)}_+(z) \\
\phi^{(\xi,\epsilon)}_-(z)
\end{array}
\right)
\]
with \cite{Erdelyi}
\[
{\cal C}^{(\xi,\epsilon)}
\; = \;
\sqrt{\frac{\epsilon}{\xi}}\left(
\begin{array}{cc}
\frac{{\rm e}^{\xi a}\Gamma(-\epsilon)\Gamma(1+\xi)}
{\Gamma\left(\frac{1+\xi+\nu-\epsilon}{2}\right)\Gamma\left(\frac{1+\xi-\nu-\epsilon}{2}\right)}
&
\frac{\Gamma(\epsilon)\Gamma(1+\xi)}
{\Gamma\left(\frac{1+\xi+\nu+\epsilon}{2}\right)\Gamma\left(\frac{1+\xi-\nu+\epsilon}{2}\right)}
\\
\frac{\Gamma(-\epsilon)\Gamma(1-\xi)}
{\Gamma\left(\frac{1-\xi+\nu-\epsilon}{2}\right)\Gamma\left(\frac{1-\xi-\nu-\epsilon}{2}\right)}
&
\frac{{\rm e}^{-\xi a}\Gamma(\epsilon)\Gamma(1-\xi)}
{\Gamma\left(\frac{1-\xi+\nu+\epsilon}{2}\right)\Gamma\left(\frac{1-\xi-\nu+\epsilon}{2}\right)}
\end{array}
\right).
\]
To take the limit $\epsilon\to 0$ we define, similarly as in (\ref{xi:to:0}),
\begin{equation}
\label{etas}
\left(
\begin{array}{c}
\phi^{(\xi)}_+(z) \\
\phi^{(\xi)}_-(z)
\end{array}
\right)
= \lim_{\epsilon\to 0}\ B_{\epsilon}
\left(
\begin{array}{c}
\phi^{(\xi,\epsilon)}_+(z) \\
\phi^{(\xi,\epsilon)}_-(z)
\end{array}
\right)
\end{equation}
and
\[
{\cal C}^{(\xi)}
\; = \;
\lim_{\epsilon\to 0}\
{\cal C}^{(\xi,\epsilon)}\cdot
B_{\epsilon}^{-1}
\]
with
\[
B_{\epsilon}
\; = \;
\frac12
\left(
\begin{array}{cc}
\sqrt{\epsilon} + \frac{1}{\sqrt\epsilon}
&
\sqrt{\epsilon} - \frac{1}{\sqrt\epsilon}
\\
\sqrt{\epsilon} - \frac{1}{\sqrt\epsilon}
&
\sqrt{\epsilon} + \frac{1}{\sqrt\epsilon}
\end{array}
\right).
\]
For $z \to x$
\[
\phi^{(\xi)}_{\pm}(z) \; = \; \sqrt{z}\left(1 \pm \frac12\log\frac{z}{x}\right)
+ o(z),
\]
and the monodromy matrix of this pair around $z=0$ (for $z \to {\rm e}^{2\pi i}z$) is
\[
M_{\phi^{(\xi)}}^{(z=0)}
\; = \;
-
\left(
\begin{array}{cc}
1+\frac{i\pi}{2} & \frac{i\pi}{2} \\
- \frac{i\pi}{2} & 1-\frac{i\pi}{2}
\end{array}
\right).
\]
One thus gets a monodromy matrix of the pair $\psi^{(\xi)}_{\pm} = \lim\limits_{\epsilon\to 0}\psi^{(\xi,\epsilon)}_{\pm}$
around $z = 0$ in the form
\begin{eqnarray*}
M_{\psi^{(\xi)}}^{(z=0)}
&   =  &
{\cal C}^{(\xi)}\cdot M_{\phi^{(\xi)}}^{(z=0)}  \cdot \left({\cal C}^{(\xi)}\right)^{-1}.
\end{eqnarray*}

The limit $\xi\to 0$ can be taken as in (\ref{xi:to:0}),
\begin{equation}
\label{psis:0}
\left(
\begin{array}{c}
\psi_{+} \\
\psi_{-}
\end{array}
\right)
\; = \;
\lim_{\xi \to 0}
B_\xi
\cdot
\left(
\begin{array}{c}
\psi^{(\xi)}_{+} \\
\psi^{(\xi)}_{-}
\end{array}
\right)
\end{equation}
with
\[
B_\xi
=
\frac12
\left(
\begin{array}{cc}
\sqrt{\xi} + \frac{1}{\sqrt\xi}
&
\sqrt{\xi} - \frac{1}{\sqrt\xi}
\\
\sqrt{\xi} - \frac{1}{\sqrt\xi}
&
\sqrt{\xi} + \frac{1}{\sqrt\xi}
\end{array}
\right).
\]
We have
\begin{eqnarray*}
\psi_+(z) & = & \sqrt{x-z}\left(1 + a\right) + \frac12\sqrt{x-z} \log\left(1-\frac{z}{x}\right) + o(z-x),
\\
\psi_-(z) & = & \sqrt{x-z}\left(1 - a\right) - \frac12\sqrt{x-z} \log\left(1-\frac{z}{x}\right) + o(z-x).
\end{eqnarray*}
The monodromy matrix of this pair around $z = x$ (in the counterclockwise direction,
$(z-x) \to {\rm e}^{2\pi i}(z-x)$)
thus reads
\[
M_{\psi}^{(z=x)}  \; = \;
-
\left(
\begin{array}{cc}
1-\frac{i\pi}{2} & -\frac{i\pi}{2} \\
\frac{i\pi}{2} & 1+\frac{i\pi}{2}
\end{array}
\right),
\]
and the monodromy matrix of the  $\psi_{\pm}$ pair around $z=0$ (in the counterclockwise direction,
$z \to {\rm e}^{2\pi i}z$) can be calculated as
\begin{eqnarray}
\label{monodr:final}
M_{\psi}^{(z=0)}
& = &
\lim_{\xi\to 0}
B_\xi
\cdot
M_{\psi^{(\xi)}}^{(z=0)}
\cdot
B_\xi
^{-1}
\\
\nonumber
& = &
\frac{i}{2\pi}
\left(
\begin{array}{cc}
2i\pi +\left(\beta^2(x)-4\right)\cos^2\frac{\pi\nu(x)}{2}
&
(\beta(x)-2)^2\cos^2\frac{\pi\nu(x)}{2}
\\
-(\beta(x)+2)^2\cos^2\frac{\pi\nu(x)}{2}
&
2i\pi- \left(\beta^2(x)-4\right)\cos^2\frac{\pi\nu(x)}{2}
\end{array}
\right).
\end{eqnarray}
Here
\[
\beta(x) \; \equiv \; \psi_0\left(\frac{1-\nu(x)}{2}\right) + \psi_0\left(\frac{1+\nu(x)}{2}\right)
+2\gamma{\scriptscriptstyle E} - 2a(x),
\]
$\psi_0(z) = \frac{d}{dz}\log\Gamma(z)$ is the digamma function and $\gamma_{\scriptscriptstyle E}$ denotes the Euler-Mascheroni constant.

Monodromy matrix around $z = 1$ can now be obtained (paying attention to the opposite orientation
of the planes $z$ and $w = 1-z$)  by repeating the calculation above
with $x$ substituted by $1-x.$  As these calculations are fairly straightforward we refrain
from writing them down explicitly.

We shall now turn to the power-like corrections. Define
\[
\left(\hat{\cal G} f\right)(z)
\; = \;
\int\limits_{x}^z\!dz'
\left(\psi_{+}(z)\psi_{-}(z')-\psi_{-}(z)\psi_{+}(z')\right) f(z')
\]
with $\psi_{\pm}$ given by (\ref{psis:0}). Since
\[
\left(\partial^2 + T_0\right)\hat{\cal G} \; = \; {\mathbf 1},
\]
we can rewrite (\ref{perturbative}) in the form of an integral equation
\[
\Psi_{\pm}
\; = \;
\psi_{\pm} +\hat{\cal G}V\,\Psi_{\pm}
\]
with a (formal) solution
\begin{equation}
\label{formal:solution}
\Psi_{\pm}
\; = \;
\left(1-\hat{\cal G}V\right)^{-1}\psi_{\pm}.
\end{equation}
From (\ref{formal:solution}) one can read off the $n-$th order correction to the functions
$\psi_{\pm}(z),$
\begin{equation}
\delta^{(n)}\psi_{\pm}(z)
 =
\int\limits_{x}^z\!\!dz_n\,{\cal G}(z,z_n)V(z_n)
\int\limits_{x}^{z_n}\!\!dz_{n-1}\, {\cal G}(z_n,z_{n-1})V(z_{n-1})
\ldots
\int\limits_{x}^{z_2}\!\!dz_{1}\, {\cal G}(z_2,z_{1})V(z_{1})\psi_{\pm}(z_1).
\end{equation}
Let us discuss in some details the case $n = 1,$
\begin{eqnarray*}
\delta^{(1)}\psi_{\pm}(z)
& = &
\psi_{+}(z)\int\limits_{x}^z\!dz'\psi_{-}(z')V(z')\psi_{\pm}(z')
-
\psi_{-}(z)\int\limits_{x}^z\!dz'\psi_{+}(z')V(z')\psi_{\pm}(z').
\end{eqnarray*}
The functions $\psi_{\pm}(z')V(z')\psi_{\pm}(z')$ have integrable (logarithmic)
singularities for $z \to x$ and $z \to 0.$ Consequently,
\begin{eqnarray*}
\delta^{(1)}\psi_{\pm}(z) & = & o\left(z-x\right)
\hskip .5cm {\rm for} \hskip .5cm z \to x,
\end{eqnarray*}
which means that the monodromy matrix of the $\psi_{\pm}$ pair remains unchanged.
For $z \to 0$ the leading correction takes the form
\begin{eqnarray*}
\delta^{(1)}\psi_{\pm}(z) & = & \alpha^{(1)}_{\pm,+}\cdot\psi_{+}(z) + \alpha^{(1)}_{\pm,-}\cdot\psi_{-}(z) + o(z),
\end{eqnarray*}
with
\[
\alpha^{(1)}_{\pm,\pm} \; = \; \int\limits_{x}^0\!dz'\psi_{\pm}(z')V(z')\psi_{\pm}(z').
\]
The monodromy matrix of the $\psi_{\pm} + \delta^{(1)}\psi_{\pm}$ pair is therefore given by
\[
M_{\psi+\delta^{(1)}\psi}^{(0)}
\; = \;
\left(1+\alpha^{(1)}\right)M_{\psi}^{(0)}\left(1+\alpha^{(1)}\right)^{-1}.
\]
Notice further that $\frac{1}{\sqrt{x}}\psi_{\pm}(x\zeta)$ does not depend on $x$ and
$V(xz) = {\cal O}\left(x^{-1}\right),$ so that (with $z' = x\zeta$)
\[
\alpha^{(1)}_{\pm,\pm}
=
x\int\limits_{1}^0\!d\zeta
\left(\frac{1}{\sqrt x}\psi_{\pm}(x\zeta)\right)
xV(x\zeta)
\left(\frac{1}{\sqrt x}\psi_{\pm}(x\zeta)\right)
= {\cal O}(x).
\]
Similarly one gets
\begin{eqnarray*}
\delta^{(n)}\psi_{\pm}(z) & = & o\left(z-x\right)
\hskip .5cm {\rm for} \hskip .5cm z \to x,
\end{eqnarray*}
and
\begin{eqnarray*}
\delta^{(n)}\psi_{\pm}(z) & = & \alpha^{(n)}_{\pm,+}\cdot\psi_{+}(z) + \alpha^{(n)}_{\pm,-}\cdot\psi_{-}(z) + o(z)
\hskip .5cm {\rm for} \hskip .5cm z \to 0,
\end{eqnarray*}
with
\[
\alpha^{(n)}_{\pm,\pm} \; = \; {\cal O}\left(x^n\right).
\]
Up to this order the monodromy of the pair $\Psi_{\pm}$  around $z = 0$ is therefore given by
\[
M_{\Psi}^{(0)} \; = \; \left(1+\sum_{k=1}^n\alpha^{(k)}\right) M_{\psi}^{(0)}
\left(1+\sum_{k=1}^n\alpha^{(k)}\right)^{-1}.
\]
Let us remark that although  the matrices $\alpha^{(k)}$ are difficult to evaluate analytically,
their numerical calculation is rather straightforward.

\section{Zamolodchikovs conjecture}
\setcounter{equation}{0}

The 4-point function of the DOZZ theory
with the operator insertions at
$z_1 = 0,$ $z_3 = 1,$ $z_4=\infty$ and $z_2 = x,$
can be expressed
as an integral of $s$-channel conformal blocks and DOZZ couplings over
the continuous spectrum of the theory.
In the semiclassical limit the integrand can be written
in terms of 3-point classical Liouville actions and the classical block \cite{Zamolodchikov:1995aa},
\begin{equation}
\label{a4}
\Big\langle
V_4(\infty,\infty)V_3(1,1)V_2(x,\bar x)V_1(0,0)
 \Big\rangle
 \sim
\int\limits_{0}^{\infty}\!dp\; {\rm e}^{-Q^2 S(\delta_i;x;\delta)}
\end{equation}
where $\delta = \frac14 + p^2$ and
\begin{equation}
\label{deltaaction}
S(\delta_i;x;\delta)
=S^{\rm (cl)}(\delta_4,\delta_3,\delta) +
S^{\rm (cl)}(\delta,\delta_2,\delta_1)- f_{\delta}
\!\left[_{\delta_{4}\;\delta_{1}}^{\delta_{3}\;\delta_{2}}\right](x)
-\bar f_{\delta}
\!\left[_{\delta_{4}\;\delta_{1}}^{\delta_{3}\;\delta_{2}}\right](\bar x).
\end{equation}
    The  3-point
classical Liouville action with a parabolic $\delta_1 = \frac14,$
an elliptic $\delta_2 =  \frac14\left(1-\xi^2\right),$ and a
hyperbolic weight $\delta =\frac14+p^2,$ reads \cite{Zamolodchikov:1995aa,Hadasz:2003he}
\begin{eqnarray}
\label{action:mixed}
\nonumber
S^{\rm ( cl)}(\delta,\delta_2,\delta_1)
& = &
-(1-\xi)\log 2 +
2F\left(\frac{1-\xi}{2} +ip\right)
+
2F\left(\frac{1-\xi}{2} -ip\right)
\\
&&-F(\xi)
 + H(2i p) + \pi|p| +{\rm const},
\end{eqnarray}
where
\[
F(x)  =  \int\limits_{1/2}^{x}\!dy\; \log\frac{\Gamma(y)}{\Gamma(1-y)},
\hskip 1cm
H(x)  =  \int\limits_0^x\!dy\;\log\frac{\Gamma(-y)}{\Gamma(y)}.
\]
$f_{\delta}
\!\left[_{\delta_{4}\;\delta_{1}}^{\delta_{3}\;\delta_{2}}\right](x)$ is the
classical conformal block \cite{Zamolodchikov:1995aa} (or
 the ``classical action'' of \cite{Zam0,Zam}), defined as the semiclassical asymptotic
\begin{equation}
\label{defccb}
{\cal F}_{\!1+6Q^2,\Delta}
\!\left[_{\Delta_{4}\;\Delta_{1}}^{\Delta_{3}\;\Delta_{2}}\right]
\!(x)
\; \sim \;
\exp \left(
Q^2\,f_{\delta}
\!\left[_{\delta_{4}\;\delta_{1}}^{\delta_{3}\;\delta_{2}}\right]
\!(x)
\right)
\end{equation}
of the BPZ conformal block \cite{Belavin:1984vu}.

In the classical limit $Q^2\to\infty$ the integral on the r.h.s.
of relation (\ref{a4}) is dominated by its saddle point value. One
thus gets
\begin{eqnarray}
\label{zc}
S^{\rm (cl)}(\delta_i;x)
&=& S(\delta_i;x;\delta_s)\\
\nonumber
&=&
S^{\rm (cl)}(\delta_4,\delta_3,\delta_s) +
S^{\rm (cl)}(\delta_s,\delta_2,\delta_1)-
f_{\delta_s}
\!\left[_{\delta_{4}\;\delta_{1}}^{\delta_{3}\;\delta_{2}}\right](x)
-\bar f_{\delta_s}
\!\left[_{\delta_{4}\;\delta_{1}}^{\delta_{3}\;\delta_{2}}\right](\bar x).
\end{eqnarray}
where
$\delta_s={\textstyle\frac{1}{4}}+p^2_s(x)$ and
the {\it saddle point momentum} $p_s(x)$ is determined by the equation
\begin{equation}
\label{saddle}
 \frac{\partial}{\partial p}
S\left(\delta_i;x;{\textstyle\frac{1}{4}}+p^2\right)_{|p=p_s}
= 0.
\end{equation}
Since the semiclassical limit should be independent of the choice of the channel in
the factorization of the DOZZ 4-point function the Zamolodchikovs conjecture (\ref{zc})
yields
three different expressions for the 4-point classical Liouville action.
The corresponding consistency equations ({\it classical bootstrap equations})
has been numerically verified for punctures \cite{Hadasz:2005gk}
and for punctures
and one and two
elliptic singularities \cite{Hadasz:200?}.

Taking into account the classical geometry corresponding
to  hyperbolic weights  \cite{sei}
one may expect
that
the saddle point momentum $p_s(x)$ in the $s$-channel is  related to the
length $\ell_s(x)$ of the closed geodesic separating the ``initial'' $z = 0,x$
from the ``final''
 $z = 1, \infty$ singularities:
\begin{equation}
\label{geodesic:length}
\ell_s(x) = 4\pi p_s(x).
\end{equation}
In the case of punctures this conjecture has strong numerical support \cite{Hadasz:2005gk}.

The
classical limit of the quantum conformal block is up to now the only method of
calculating the classical conformal block\footnote{
Let us note that it is by no means obvious that
the quantum conformal block for "heavy" weights does have the asymptotic of the form
assumed in (\ref{defccb}). This is however very well supported
by symbolic calculations for
first few terms in a number of cases.}.
An efficient recursive
method of calculating coefficients of the  expansion of the
quantum block in powers of $x$
 were  developed by Al.\ Zamolodchikov \cite{Zam0}.
Using this method and taking the limit $Q\to \infty$ one can
 calculate term by term the coefficients
of  the power expansion
\begin{equation}
\label{classblock}
f_{\delta}\!\left[_{\delta_{4}\;\delta_{1}}^{\delta_{3}\;\delta_{2}}\right]\!(\,x)
= (\delta-\delta_1-\delta_2) \log x +  \sum_{n=1}^\infty
x^{\;n} f^{\,n}_{\delta}\!\left[_{\delta_{4}\;\delta_{1}}^{\delta_{3}\;\delta_{2}}\right]\ .
\end{equation}
 In the
case of three punctures and one elliptic singularity  the first few terms of
this expansion read
\begin{eqnarray}
\label{classblock:x}
f_{\frac14 + p^2}
\!\left[\;_{{1\over 4}\;\;\;\;\;\, {1\over 4}}^{{1\over 4}\;\;\;{1-\xi^2\over 4}}\;\right]
\!(x)
& = &
\left(p^2 -{1-\xi^2\over 4}\right)\log x
+
\left({1-\xi^2\over 8}+{p^2\over 2}\right)x
\\
\nonumber
& + &
\left( {9\left(1-\xi^2\right)\over 128} + {13\ p^2\over 64} + {\left(1-\xi^2\right)^2\over 1024\ (1 + p^2)}\right)x^2
+{\cal O}\left(x^3\right).
\end{eqnarray}

The limitation of  formulae (\ref{classblock}) and (\ref{classblock:x}) is that the power series
involved are supposed to converge only for $|x|<1$. A more convenient representation of the
conformal block
was developed  by Al.\ Za\-mo\-lod\-chi\-kov  in \cite{Zam}, where he proposed to regard the block as a function of
the variable
$$
q(x) = {\rm e}^{-\pi {K(1-x)\over K(x)}},
\hskip 1cm
K(x)=\int\limits_{0}^1{dt\over \sqrt{(1-t^2)(1-xt^2)}}.
$$
In terms of  $q$  the classical
conformal block reads:
\begin{eqnarray}
\label{Hclassblock}
f_{\delta}\!\left[_{\delta_{4}\;\delta_{1}}^{\delta_{3}\;\delta_{2}}\right]\!(\,x)
&=&
 ({\textstyle {1\over 4}}-\delta_1-\delta_2) \log x
+
 ({\textstyle {1\over 4}}-\delta_2-\delta_3) \log (1-x)\\
 \nonumber
&+&
 ({\textstyle {1\over 4}}-2(\delta_1+\delta_2+\delta_3+\delta_4)) \log
 \left( {\textstyle{2\over\pi}}K(x)\right)\\
 \nonumber
 &+& (\delta -{\textstyle {1\over 4}}) \log 16
 - (\delta -{\textstyle {1\over 4}}) \pi {K(1-x)\over K(x)}+
  h_{\delta}\!\left[_{\delta_{4}\;\delta_{1}}^{\delta_{3}\;\delta_{2}}\right](q),
\end{eqnarray}
where
\begin{equation}
\label{hexp}
 h_{\delta}\!\left[_{\delta_{4}\;\delta_{1}}^{\delta_{3}\;\delta_{2}}\right](q)
 =\sum_{n=1}^\infty
(16q)^{n} h^{\,n}_{\delta}\!\left[_{\delta_{4}\;\delta_{1}}^{\delta_{3}\;\delta_{2}}\right]
\end{equation}
is supposed to converge  uniformly
on each subset $\{q:|q|<{\rm e}^{-\epsilon}<1\}$.

The coefficients of  power series (\ref{hexp}) can be determined term by term
using Zamolodchikovs recursive method to calculate coefficients in the
$q$-expansion of
the quantum block \cite{Zam} and then taking the limit $Q\to \infty$.
This for instance  yields:
\begin{eqnarray}
\label{classblock:q}
h_{{1\over 4} +p^2 }\!\left[\;_{{1\over 4}\;\;\;\;\;\, {1\over 4}}^{{1\over 4}\;\;\;{1-\xi^2\over 4}}\;\right]\!(\,q)
& = &
  \frac{\left(1-\xi^2\right)^2}{4\left( 1 + p^2 \right) }\,q^2
 +\left[
 \frac{\left(1-\xi^2\right)^2\left(15+18\xi^2-\xi^4\right)}{128(1+p^2)}
\right.
\\
\nonumber
&+ &
\left.
 \frac{\left(1-\xi^2\right)^2\left(9-\xi^2\right)^2}{128(4+p^2)}
 +
 \frac{3\left(1-\xi^2\right)^4}{128\left(1+p^2\right)^2}
 -
 \frac{\left(1-\xi^2\right)^4}{32\left(1+p^2\right)^3}
 \right]q^4
+{\cal O}\left(q^6\right).
\end{eqnarray}

In the case of four punctures  the
saddle point equation determining $p_s(x)$ reads
\begin{equation}
\label{saddle:p}
0 =
 \frac{\partial}{\partial p}
S\left({\textstyle
\frac{1}{4};x;\frac{1}{4}+p^2}\right) = 2\pi +
4i\log\frac{\Gamma^2\left(\frac12 +
ip\right)\Gamma(-2ip)}{\Gamma^2\left(\frac12 -
ip\right)\Gamma(2ip)} -2\Re\,\frac{\partial}{\partial
p}\,f_{\frac14+p^2}
\!\left[_{\frac14\;\frac14}^{\frac14\;\frac14}\right](x),
\end{equation}
and the classical Liouville action is given by
\begin{equation}
\label{classical:fin}
{\cal S}^{\rm (cl)}\left({\textstyle \frac14};x\right)
\; = \;
S\left({\textstyle \frac14; x;\frac14 + p^2_s(x)}\right).
\end{equation}
Using (\ref{deltaaction}),
(\ref{action:mixed})  and (\ref{classblock:x}) one gets:
\begin{eqnarray}
\label{c2}
\nonumber
c_2(x)
& = &
-\frac{\partial {\cal S}^{(\rm cl)}(\frac14;x)}{\partial x}
\; = \;
- \left.\frac{\partial S\left(\frac14;x;\frac14 + p^2\right)}{\partial p}\right|_{p = p_s(x)}
\hspace*{-5mm}\cdot\frac{\partial p_s(x)}{\partial x}
- \left.\frac{\partial S\left(\frac14;x;\frac14 + p^2\right)}{\partial x}\right|_{p = p_s(x)}
\\
& = &
- \left.\frac{\partial S\left(\frac14;x;\frac14 + p^2\right)}{\partial x}\right|_{p = p_s(x)}
\; = \;
\left.\frac{\partial}{\partial x}\,f_{\frac14+p^2}
\!\left[_{\frac14\;\frac14}^{\frac14\;\frac14}\right](x)\right|_{p = p_s(x)}
\\
\nonumber
& = &
\frac{4p_s^2(x) -1}{4x}
+\frac18\left(4p_s^2(x)  +1\right)  +
\left[\frac92 +13p_s^2(x) +\frac18\frac{1}{1+p_s^2(x)}\right]\frac{x}{32}
+ {\cal O}\left(x^2\right)
\end{eqnarray}
or, employing the $q$ expansion of the classical block,
\begin{eqnarray}
\label{c2:q}
\nonumber
&&
\hspace*{-10mm}
c_2(x)
=
-\frac{1}{4x(1-x)}\left[1+\frac74\left(\frac{E(x)}{K(x)} + x-1\right)\right]
-
\frac{\pi^2}{4x(1-x)K^2(x)}\left[
p_s^2(x) +
\frac{1}{1+p_s^2(x)}\frac{q^2}{2}
\right.
\\
\nonumber
&&
\\
&&
\left.
+
\left(
\frac{15}{1+p_s^2(x)}
+
\frac{81}{4+p_s^2(x)}
+
\frac{3}{\left(1+p_s^2(x)\right)^2}
-
\frac{4}{\left(1+p_s^2(x)\right)^3}
\right)\frac{q^4}{32}
+ {\cal O}\left(q^6\right)
\right],
\end{eqnarray}
where $E(x)$ denotes the complete elliptic integral of the second kind.
Equations (\ref{c2}), (\ref{c2:q}) are simple consequences of the Zamolodchikovs
conjecture. They provide a new relation between the accessory parameters,
the classical
conformal block  and the geodesic length function $\ell_s(x)$ which is certainly worth
further investigations.

In order to calculate  $a_2 \equiv a$ for four punctures one can replace one
puncture by an elliptic singularity and then take the limit
\[
2a \; = \; \lim_{\xi\to 0}\frac{1}{\xi}\log A^2.
\]
Using (\ref{A:j}), (\ref{deltaaction}), (\ref{action:mixed}) and (\ref{classblock:x})
one obtains
{\small
\begin{eqnarray}
\label{a}
\nonumber
&&
2a(x) \; = \;
\lim_{\xi\to 0}
\left\{
\frac{1}{\xi}
\log\frac{\Gamma\left(\frac{1+\xi}{2}-ip_s(x)\right)\Gamma\left(\frac{1+\xi}{2}+ip_s(x)\right)}
{\Gamma\left(\frac{1-\xi}{2}-ip_s(x)\right)\Gamma\left(\frac{1-\xi}{2}+ip_s(x)\right)}
-\frac{2}{\xi}
\Re\,\frac{\partial}{\partial\xi}\left.
f_{\frac14+p^2}\!\left[^{\frac14\,\frac{1-\xi^2}{4}}_{\frac14\ \;\;\frac14}\right]\!(x)
\right|_{p = p_s(x)}\right\}
\\
&& = \;
\psi_0\left(\frac12 + ip_s(x)\right) + \psi_0\left(\frac12 - ip_s(x)\right)
- \frac{2}{\xi}
\Re\,\frac{\partial}{\partial\xi}\left.
f_{\frac14+p^2}\!\left[^{\frac14\,\frac{1-\xi^2}{4}}_{\frac14\ \;\;\frac14}\right]\!(x)
\right|_{p = p_s(x),\xi = 0}
\\
\nonumber
&& \\
\nonumber
&&= \;
\psi_0\left(\frac12 + ip_s(x)\right) + \psi_0\left(\frac12 - ip_s(x)\right)
 -\frac12\log x\bar x +\frac{\Re\,x}{2} +\left[36+\frac{1}{1+p_s^2(x)}\right]\frac{\Re\,x^2}{2^7}
+ {\cal O}\left(x^3\right),
\end{eqnarray}
}
or, using (\ref{Hclassblock}) and (\ref{classblock:q}):
\begin{eqnarray}
\label{a:q}
\nonumber
&&
\hspace*{-5mm}
2a(x)
\; = \;
\psi_0\left(\frac12 + ip_s(x)\right) + \psi_0\left(\frac12 - ip_s(x)\right)
\\
&&
-\; \frac12\log x\bar x - \frac12\log\left|1-x\right|^2 - \log\left|\frac{2}{\pi}K(x)\right|^2
+ \frac{2}{1+p_s^2(x)} \Re\,q^2
\\
\nonumber
&&
+ \;
\left[
\frac{45}{4+p_s^2(x)}
+
\frac{3}{1+p_s^2(x)}
+
\frac{3}{\left(1+p_s^2(x)\right)^2}
-
\frac{4}{\left(1+p_s^2(x)\right)^3}
\right]
\frac{\Re\,q^4}{8}
+{\cal O}\left(q^6\right).
\end{eqnarray}
Note that in the formulae above one only needs the saddle point momentum $p_s(x)$
for four punctures ($\xi = 0$).

Let us finally turn to the problem of determining the saddle point momentum
$p_s(x).$ As was discussed in \cite{Hadasz:2005gk} $p_s(x)$ can be determined numerically,
using the $q$ expansion of the classical block,
with an essentially arbitrary high precision everywhere but at
small vicinities of the singular points $x =1$ and $x = \infty.$
On the other hand, the problem of analytic determination
of the saddle point momentum still remains to be solved and only partial results are available.

Both  geometrical arguments and the form of (\ref{classblock:x}) indicate that
for $x \to 0$ the solution of (\ref{saddle:p}) should also tend to zero. For $p\to 0$
\begin{equation}
\label{expansion:small:p}
4i\log\frac{\Gamma^2\left(\frac12 + ip\right)\Gamma(-2ip)}{\Gamma^2\left(\frac12 - ip\right)\Gamma(2ip)}
= -4\pi + 32 p \log 2
+16\sum\limits_{k=1}^\infty
\frac{(-1)^k\zeta(2k+1)}{2k+1}
\left(2^{2k}-1\right)p^{2k+1}
\end{equation}
so that, up to the leading terms, the saddle point equation (\ref{saddle:p}) takes the form
\begin{equation}
\label{leading:eq}
-\pi + 16p\log 2 - p\log x \bar x = 0
\end{equation}
and, to this order,
\begin{equation}
\label{leading:sol}
p_s(x) \sim \varepsilon(x) \equiv \frac{\pi}{-\log x\bar x + 16\log 2}.
\end{equation}
We can now solve (\ref{saddle:p}) by iteration, in the
form of a double series expansion in $\Re x$ and $\varepsilon(x)$
(since $x = {\rm exp}\left(-\frac{1}{1/(-\log x)}\right),$
the powers of $x$ can be viewed as ``non--perturbative'' corrections
to the $\varepsilon(x)$ series).
For instance, keeping in (\ref{saddle:p})
terms up to $p^3$ and $x^2$ we get
\begin{eqnarray}
\label{p:s}
\nonumber
p_s & = &
\frac{\pi}{-\log x\bar x + 16\log 2 -\Re\,x - \frac{207}{512}\Re\,x^2}
+ \\
\nonumber
&&
\frac{8\zeta(3) +\frac{1}{256}\Re\,x^2}
{-\log x\bar x + 16\log 2 -\Re\,x - \frac{207}{512}\Re\,x^2}\ p_s^3
+{\cal O}\left(\varepsilon(x)^6, x^3\varepsilon(x)\right)
\\
\nonumber
& = &
\frac{\pi}{-\log x\bar x + 16\log 2 -\Re\,x - \frac{207}{512}\Re\,x^2}
+ \\
&&
\frac{8\pi^3\zeta(3) +\frac{\pi^3}{256}\Re\,x^2}
{\left(-\log x\bar x + 16\log 2 -\Re\,x - \frac{207}{512}\Re\,x^2\right)^4 }
+{\cal O}\left(\varepsilon(x)^6, x^3\varepsilon(x)\right)
\\
\nonumber
& = & \varepsilon(x) + \frac{8\zeta(3)}{\pi}\varepsilon^4(x)
+
\left(\frac{1}{\pi}\varepsilon^2(x) +\frac{32\zeta(3)}{\pi^2}\varepsilon^5(x)\right)\Re\,x +
\frac{\varepsilon^2(x)}{\pi^2}\left(\Re\,x\right)^2 +
\\
\nonumber
&&
\left(\frac{207}{512\pi}\varepsilon^2(x) + \frac{1}{256\pi}\varepsilon^4(x)
+
\frac{207\zeta(3)}{16\pi^2}\varepsilon^5(x)\right)\Re\,x^2
+{\cal O}\left(\varepsilon(x)^6, x^3\varepsilon(x)\right).
\end{eqnarray}
As in the case of the accessory parameters, one can also work out
the formula for the saddle point momentum involving the $q$ expansion
of the classical conformal block.

For a sufficiently small $x$ the r.h.s. of (\ref{p:s}) agrees with the numerically
calculated saddle point momentum and is well within the know analytic bounds
on $\frac{\ell_s(x)}{4\pi}$ \cite{Hadasz:2005gk,Hempel}. However, to determine
the radius of convergence of the series in (\ref{p:s}) or to give an estimate
on the omitted terms one would need to know the classical conformal block exactly.

As it was discussed in the previous section in order to determine
the monodromy matrix at $x=1$ one needs a power expansion of
$p_s(x)$ at this point. In other words, an analytic continuation of
the classical block from the vicinity of $x =0$ to $x =1$ is
required. One possible approach to this problem is to consider the
classical limit of the braiding relation for the quantum BPZ block
(for conformal blocks corresponding to degenerated fields this
calculation is quite straightforward).

We believe that a better understanding of the
classical conformal block and the geodesic length function
may provide an essentially new insight into the
problem of finding an analytic expressions for the map $\rho$ and the uniformizing
group.
Further studies of these structures are definitely worth pursuing.

\section*{Acknowledgments}
This work was partially supported by the Polish State Research Committee (KBN) grant
no.\ 1 P03B 025 28.

\vskip 1mm
\noindent
The research of L.H. is supported by the Alexander von Humboldt Foundation scholarship.

\vskip 1mm
\noindent
Part of the ideas contained in this work emerged
during the 2005 Simons Workshop in Physics and Mathematics.
L.H.\  thanks the organizers of the Workshop for their warm hospitality,
which created at Stony Brook a highly stimulating environment
and for the financial support.

\end{document}